\documentstyle[preprint,prl,aps]{revtex}

\begin{document}

\preprint{\vbox{\hbox {February 1999} \hbox{IFP-768-UNC} \hbox{hep-th/9902168}}}

\title{\bf Conformal ${\cal N}=0~~d=4$ Gauge Theories from AdS/CFT Superstring Duality?}
\author{\bf Paul H. Frampton and William F. Shively}
\address{Department of Physics and Astronomy,}
\address{University of North Carolina, Chapel Hill, NC  27599-3255}
\maketitle

\begin{abstract}
Non-supersymmetric $d=4$ gauge theories which arise from superstring duality
on a manifold $AdS_5 \times S_5/Z_p$ are cataloged for a range
$2 \leq p \leq 41$.
A number have vanishing two-loop gauge $\beta-$function,
a necessary but not sufficient condition to be a conformal field theory.  
\end{abstract}


\newpage

\bigskip
\bigskip

The relationship of the Type IIB superstring to conformal gauge theory
in $d=4$ gives rise to an interesting class of gauge 
theories\cite{Maldacena,Gubser,Witten,Susskind}.
Choosing the simplest compactification\cite{Maldacena}
on $AdS_5 \times S_5$ gives rise to an ${\cal N} = 4$ SU(N) gauge theory
which has been known for some time\cite{Mandelstam} to be conformal due to
the extended global supersymmetry and non-renormalization theorems. All
of the RGE $\beta-$functions for this ${\cal N} = 4$ 
case are vanishing in perturbation theory.

One of us (PHF) has recently\cite{Frampton} pursued the idea
that an ${\cal N} = 0$ theory, without spacetime supersymmetry, arising
from 
compactification
\cite{Kachru,Ferrara,Russo,Schmaltz,Berkooz,Distler,Harvey,Klebanov,Kakushadze,Tsyetlin} 
on the orbifold $AdS_5 \times S_5/\Gamma$ (with $\Gamma \not\subset SU(3)$)
could be conformal and, further, could accommodate the standard model.
In the present note we systematically catalog the available
${\cal N}=0$ theories for $\Gamma$ an abelian discrete group $\Gamma=Z_p$.
We also find the subset which has $\beta_g^{(2)}=0$, a vanishing two-loop
$\beta-$function for the gauge coupling, according to the criteria of \cite{Frampton}.
In a future publication, we hope to find
how many if any of the surviving theories satisfy $\beta_Y^{(2)}=0$
and $\beta_H^{(2)}=0$ for the Yukawa and Higgs self-coupling two-loop
RGE $\beta-$functions respectively. Note that the one-loop $\beta-$functions
satisfy $\beta_Y^{(1)}=0$
and $\beta_H^{(1)}=0$ because they are leading order in the planar
expansion\cite{Vafa,Bershadsky,Lawrence,Shatashvili}. All one-loop
${\cal N} = 0$ calculations coincide with those of the 
conformal ${\cal N} = 4$ theory to leading order in $1/N$. However, beyond
large $N$ and beyond one-loop this coincidence
ceases, in general.

The ideas in Frampton\cite{Frampton}
concerning the cosmological constant
and model building beyond the standard model 
provide the motivation as follows. At a scale
sufficiently above the weak scale the masses and VEVs of the standard model
obviously become negligible.
Consider now that the standard model is promoted by additional states
to a conformal theory of the $d = 4$ ${\cal N} = 0$ type
which will be highly constrained or even unique, 
as well as scale invariant.
Low energy masses and VEVs are introduced softly into this conformal theory
such as to preserve the desirable properties
of vanishing 
vacuum energy and hence vanishing cosmological constant.
Since no supersymmetry breaking is needed and provided the introduction
of scales is sufficiently mild it is expected that a zero
cosmological constant can be retained in this approach. 

\bigskip

The embedding of $\Gamma = Z_p$ in the complex three-dimensional space
${\cal C}_3$ can be conveniently specified by three integers
$a_i = (a_1, a_2, a_3)$. The action of $Z_p$ on the three complex
coordinates $(X_1, X_2, X_3)$ is then:

\begin{equation}
(X_1, X_2, X_3)~~~\stackrel{Z_p}{\longrightarrow}~~~(\alpha^{a_1}X_1, \alpha^{a_2}X_2, \alpha^{a_3}X_3)
\end{equation}
\noindent where $\alpha = exp(2\pi i/p)$ and the elements of $Z_p$ are $\alpha^r$ ($0 \leq r \leq (p - 1)$).

The general rule for 
breaking supersymmetries is that for $\Gamma \subset SU(2)$,
there remains ${\cal N} = 2$ supersymmetry; $\Gamma \subset SU(3)$ leaves
${\cal N} = 1$ supersymmetry; and for $\Gamma \not\subset SU(3)$,
no supersymmetry (${\cal N} = 0$) survives.

To ensure that $\Gamma \not\subset SU(3)$ the requirement is that

\begin{equation}
a_1~+~a_2~+~a_3~\neq~0~~(mod~~p)
\label{modp}
\end{equation}
Each $a_i$ can, without loss of generality, be in the range
$0 \leq a_i \leq (p - 1)$. Further we may set $a_1 \leq a_2 \leq a_3$
since permutations of the $a_i$ are equivalent.  Let us define
$\nu_k(p)$ to be the number of possible ${\cal N} = 0$ theories with
$k$ non-zero $a_i$ ($1 \leq k \leq 3$). 

Since $a_i = (0, 0, a_3)$
is clearly equivalent to $a_i = (0, 0, p - a_3)$ the value of
$\nu_1(p)$ is

\begin{equation}
\nu_1(p) = \lfloor p/2 \rfloor 
\label{nu1}
\end{equation}
where $\lfloor x \rfloor$ is the largest integer not greater than $x$. 

For $\nu_2(p)$ we observe that $a_i = (0, a_2, a_3)$ is equivalent to
$a_i = (0, p - a_3, p - a_2)$. Then we may derive, taking into
account Eq.(\ref{modp}) that, for $p$ {\it even}

\begin{equation}
\nu_2 (p)  = 2\sum_{r = 1}^{\lfloor \frac{p-2}{2} \rfloor} r = \frac{1}{4} p(p - 2)
\end{equation}
while, for $p$ {\it odd}

\begin{equation}
\nu_2(p) = 2\sum_{r=1}^{\lfloor \frac{p - 2}{2} \rfloor} r  + \lfloor \frac{p}{2} \rfloor  =  
 \frac{1}{4} (p - 1)^2
\label{nu2}
\end{equation}

For $\nu_3(p)$, the counting is only slightly more intricate.
There is the equivalence of $a_i = (a_1, a_2, a_3)$ with
$(p-a_3, p-a_2, p-a_1)$ as well as Eq.(\ref{modp}) to contend with.

In particular the theory $a_i = (a_1, p/2, p-a_1)$ is a self-equivalent (SE)
one;
let the number of such theories be $\nu_{SE}(p)$. Then
it can be seen that $\nu_{SE}(p) = p/2$ for p even, and $\nu_{SE}(p) = 0$ 
for p odd. With regard to Eq.(\ref{modp}), let $\nu_p(p)$ be the
number of theories with $\sum a_i = p$ and $\nu_{2p}(p)$ be the number
with $\sum a_i = 2p$. Then because of the equivalence of
$(a_1, a_2, a_3)$ with $(p-a_3, p-a_2, p-a_1)$, it follows that
$\nu_p(p) = \nu_{2p}(p)$. The value will be calculated below; in terms of
it $\nu_3(p)$ is given by
\begin{equation}
\nu_3(p) = \frac{1}{2} [ \bar{\nu}(p) - 2\nu_p(p) + \nu_{SE}(p)] 
\label{nu3}
\end{equation}

\noindent where $\bar{\nu}(p)$ is the number of unrestricted $(a_1, a_2, a_3)$
satisfying  $1 \leq a_i \leq (p-1)$ and
$a_1 \leq a_2 \leq a_3$. Its value is given by

\begin{equation}
\bar{\nu}(p) = \sum_{a_3=1}^{p-1} \sum_{a_3=1}^{p-1} a_2 = \frac{1}{6} p (p^2 - 1)
\end{equation}
It remains only to calculate $\nu_p(p)$ given by

\begin{equation}
\nu_p(p) = \sum_{a_1=1}^{\lfloor \frac{p}{3} \rfloor} 
\left( \left\lfloor \frac{p-a_1}{2} \right\rfloor - a_1 +1 \right)
\end{equation}
The value of $\nu_p(p)$ depends on the remainder when $p$ is divided by 6.
To show one case in detail. consider $p = 6k$ where $k$ is
an integer. Then

\begin{equation}
\nu_p(p) = \sum_{a_1 = odd}^{2k-1} \left(3k + \frac{1}{2} -
\frac{3a_1}{2} \right) + \sum_{a_1 = even}^{2k} 
\left( 3k + \frac{3a_1}{2} + 1 \right) = 3k^2 = \frac{1}{12}p^2
\end{equation}
Hence from Eq.(\ref{nu3})

\begin{equation}
\nu_3(p) = \frac{1}{2} \left[ \frac{1}{6}p(p^2-1) - 
\frac{1}{6}p^2 + \frac{p}{2} \right] 
= \frac{p}{12} (p^2 - p + 2)
\end{equation}
Taking $\nu_1(p)$ from Eq.(\ref{nu1}) and $\nu_2(p)$ from Eq.(\ref{nu2})
we find for $p = 6k$

\begin{equation}
\nu_{TOTAL}(p) = \nu_1(p) + \nu_2(p) + \nu_3(p) = \frac{p}{12}(p^2+2p+2)
\end{equation}
For $p = 6k+1$ or $p=6k+5$ one finds similarly

\begin{equation}
\nu_3(p) = \frac{1}{12}(p-1)^2(p+1)~~~~(p=6k+1~~or~~6k+5)
\end{equation}
\begin{equation}
\nu_{TOTAL} = \frac{1}{12}(p-1)(p+1)(p+2)~~~~(p=6k+1~~or~~6k+5)
\end{equation}
For $p=6k+2$ or $p=6k+4$
\begin{equation}
\nu_3(p) = \frac{1}{12}(p+1)(p^2-2p+4)~~~~(p=6k+2~~or~~6k+4)
\end{equation}
\begin{equation}
\nu_{TOTAL} = \frac{1}{12}(p^3 + 2p^2 + 2p +4)~~~~(p=6k+2~~or~~6k+4)
\end{equation}
and finally for $p=6k+3$
\begin{equation}
\nu_3(p) = \frac{1}{12}(p^3-p^2-p-3)~~~~(p=6k+3)
\end{equation}
\begin{equation}
\nu_{TOTAL} =  \frac{1}{12}(p^3 + 2p^2 -p -6)~~~~(p=6k+3)
\end{equation}
The values of $\nu_1(p)$, $\nu_2(p)$, $\nu_3(p)$, $\nu_{TOTAL}(p)$ 
and $\sum_{p'=2}^{p} \nu_{TOTAL}(p')$
for $2 \leq p \leq 41$ are listed in Table 1.

The next question is: of all these candidates for conformal ${\cal N} = 0$
theories, how many if any are conformal? As a first sifting
we can apply the criterion found in\cite{Frampton} from 
vanishing of the two-loop RGE $\beta-$function 
$\beta_g^{(2)}=0$, for the gauge coupling.
The criterion 
is that $a_1 + a_2 = a_3$. Let us denote the number of theories
fulfilling this by $\nu_{alive}(p)$.

If p is odd there is no contamination by self-equivalent possibilities
and the result is

\begin{equation}
\nu_{alive} = \sum_{r=1}^{\frac{p-1}{2}} (p-2r) = \frac{1}{4} (p-1)^2 ~~~~ (p = odd)
\label{alive}
\end{equation}
For p even some self equivalent cases must be subtracted. The sum in
Eq. (\ref{alive}) is $\frac{1}{4}p(p-2)$ and the number of self-equivalent
cases to remove is $\lfloor p/4 \rfloor$ with the results

\begin{equation}
\nu_{alive} = \frac{1}{4} p(p-3)~~~~(p = 4k)
\end{equation}
\begin{equation}
\nu_{alive} = \frac{1}{4} (p-1)(p-2)~~~~(p = 4k+2)
\end{equation}
In the last two columns of Table 1 are the values of $\nu_{alive}(p)$
and $\sum_{p'=2}^{p} \nu_{alive}(p')$.

Asymptotically for large p the ratio $\nu_{alive}(p)/\nu_{TOTAL}(p) \sim 3/p$
and hence vanishes although $\nu_{alive}(p)$ diverges; the value of the ratio 
is {\it e.g.} 0.28 at p = 5 and at p = 41 is 0.066. It is being studied 
how the two-loop requirements $\beta_Y^{(2)}=0$ and 
$\beta_H^{(2)}=0$ select from such theories. That result will further indicate
whether any $\nu_{alive}(p)$ can survive to all orders. 

\bigskip
\bigskip
\bigskip

One of us (PHF) thanks David Morrison for a discussion. This work
was supported in part by the US Department of Energy
under Grant No. DE-FG02-97ER-41036.

\newpage

Table 1. Values of $\nu_1(p)$, $\nu_2(p)$, $\nu_3(p)$, $\nu_{TOTAL}(p)$, 
$\sum_{p'=1}^{p} \nu_{TOTAL}(p')$, $\nu_{alive}(p)$ 
and $\sum_{p'=2}^p \nu_{alive}(p')$ for $2 \leq p \leq 41$.

\bigskip

\[ \begin{array}{||c||ccc|c|c||c|c||} \hline
p & \nu_1(p) & \nu_2(p) & \nu_3(p) & \nu_{TOTAL}(p) & \sum \nu_{TOTAL} & \nu_{alive}(p) &
\sum \nu_{alive}(p) \\ \hline
2 & 1 & 0 & 1 & 2 & 2 & 0 & 0 \\
3 & 1 & 1 & 1 & 3 & 5 & 1 & 1 \\
4 & 2 & 2 & 5 & 9 & 14 & 1 & 2 \\ 
5 & 2 & 4 & 8 & 14 & 28 & 4 & 6 \\
6 & 3 & 6 & 16 & 25 & 53 & 5 & 11 \\
7 & 3 & 9 & 24 & 36 & 89 & 9 & 20 \\
8 & 4 & 12 & 39 & 55 & 144 & 10 & 30 \\
9 & 4 & 16 & 53 & 73 & 217 & 16 & 46 \\
10 & 5 & 20 & 77 & 102 & 319 & 18 & 64 \\ \hline
11 & 5 & 25 & 100 & 130 & 449 & 25 & 89 \\
12 & 6 & 30 & 134 & 170 & 619 & 27 & 116 \\
13 & 6 & 36 & 168 & 210 & 829 & 36 & 152 \\
14 & 7 & 42 & 215 & 264 & 1093 & 39 & 191 \\
15 & 7 & 49 & 261 & 317 & 1410 & 49 & 240 \\
16 & 8 & 56 & 323 & 387 & 1797 & 52 & 292 \\
17 & 8 & 64 & 384 & 456 & 2253 & 64 & 356 \\
18 & 9 & 72 & 462 & 543 & 2796 & 68 & 424 \\
19 & 9 & 81 & 540 & 630 & 3426 & 81 & 505 \\
20 & 10 & 90 & 637 & 737 & 4163 & 85 & 590 \\ \hline
\end{array} \]

\newpage

Table 1 (continued)

\bigskip

\[ \begin{array}{||c||ccc|c|c||c|c||} \hline
p & \nu_1(p) & \nu_2(p) & \nu_3(p) & \nu_{TOTAL}(p) & \sum \nu_{TOTAL} & \nu_{alive}(p) &
\sum \nu_{alive}(p) \\ \hline
21 & 10 & 100 & 733 & 843 & 5006 & 100 & 690 \\
22 & 11 & 110 & 851 & 972 & 5978 & 105 & 795 \\
23 & 11 & 121 & 968 & 1100 & 7078 & 121 & 916 \\
24 & 12 & 132 & 1108 & 1252 & 8330 & 126 & 1042 \\
25 & 12 & 144 & 1248 & 1404 & 9734 & 144 & 1186 \\  
26 & 13 & 156 & 1413 & 1582 & 11316 & 150 & 1336 \\
27 & 13 & 169 & 1577 & 1759 & 13075 & 169 & 1505 \\
28 & 14 & 182 & 1769 & 1965 & 15040 & 175 & 1680 \\
29 & 14 & 196 & 1960 & 2170 & 17210 & 196 & 1876 \\
30 & 15 & 210 & 2180 & 2405 & 19615 & 203 & 2079 \\ \hline
31 & 15 & 225 & 2400 & 2640 & 22255 & 225 & 2304 \\ 
32 & 16 & 240 & 2651 & 2907 & 25162 & 232 & 2536 \\
33 & 16 & 256 & 2901 & 3173 & 28335 & 256 & 2792 \\
34 & 17 & 272 & 3185 & 3474 & 31809 & 264 & 3056 \\
35 & 17 & 289 & 3468 & 3774 & 35583 & 289 & 3345 \\
36 & 18 & 306 & 3796 & 4110 & 39693 & 297 & 3642 \\
37 & 18 & 324 & 4104 & 4446 & 44139 & 324 & 3966 \\
38 & 19 & 342 & 4459 & 4820 & 48959 & 333 & 4299 \\
39 & 19 & 361 & 4813 & 5193 & 54152 & 361 & 4660 \\
40 & 20 & 380 & 5207 & 5607 & 59759 & 370 & 5030 \\ \hline
41 & 20 & 400 & 5600 & 6020 & 65779 & 400 & 5430 \\ \hline
\end{array} \]

\end{document}